\documentclass[twoside,slac_one]{revtex4}
\usepackage{graphicx}
\usepackage{fancyhdr}
\usepackage{amsmath} 
\usepackage{bm}
\usepackage{amsxtra}
\usepackage{amssymb}
\usepackage{amsthm}
\usepackage{latexsym}
\usepackage{lscape}

\begin{document}

\title{Cosmic Ray in the Northern Hemisphere: Results from the Telescope
Array Experiment}

%

\author{C.~C.~H.~Jui$^{\dagger}$, for the Telescope Array Collaboration}
\affiliation{$^{\dagger}$Department of Physics and Astronomy, University of Utah,
Salt Lake City, UT, USA}

\begin{abstract} The Telescope Array (TA) is the largest ultrahigh energy
(UHE) cosmic ray observatory in the northern hemisphere TA is a hybrid
experiment with a unique combination of fluorescence detectors and a
stand-alone surface array of scintillation counters.  We will present the
spectrum measured by the surface array alone, along with those measured by
the fluorescence detectors in monocular, hybrid, and stereo mode.  The
composition results from stereo TA data will be discussed.  Our report will
also include results from the search for correlations between the pointing
directions of cosmic rays, seen by the TA surface array, with active
galactic nuclei. 
\end{abstract}

\maketitle

\thispagestyle{fancy}


\section{Introduction}

The Telescope Array (TA) experiment \cite{dpf2011_cchjui:TA} is a collaboration of
26 universities and research institutions in Japan, U.S., South Korea,
Russia, and Belgium.  The core of the collaboration consists of key members
from the Akeno Giant Air Shower Array (AGASA) \cite{dpf2011_cchjui:agasa} in
Japan, and the High Resolution Fly's Eye (HiRes) experiment
\cite{dpf2011_cchjui:hires}.  TA combines the large area scintillation ground
array technique developed by AGASA with that of the air fluorescence method
pioneered by the Fly's Eye (FE) \cite{dpf2011_cchjui:fe} at the University of
Utah, and later improved by the HiRes group.

\begin{figure}[ht]
\centerline{
  \includegraphics[width=80mm]{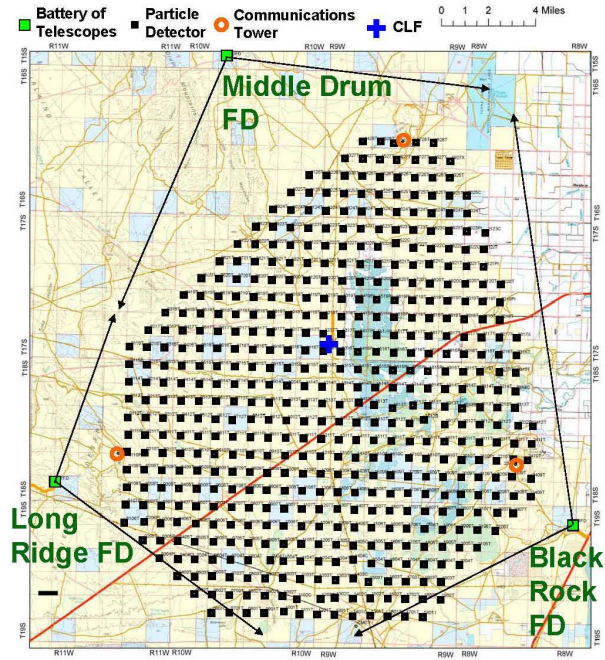}
}
\caption{The layout of the Telescope Array experiment.  The black squares
show the location of the 507 SD's.  The green squares mark the
FD stations at the periphery of the ground array. The three communications
towers near the FD stations are indicated by orange circles.  The central
laser facility, at the center of the array, is shown by the blue cross.}
\label{dpf2011_cchjui:fig001}
\end{figure}

Telescope Array is located in the central western desert of Utah, near the
city of Delta, about 250~km south west of Salt Lake City.  The arrangement of
the experiment is shown in figure~\ref{dpf2011_cchjui:fig001}.  The new experiment
consists of three fluorescence detector (FD) stations, marked in the figure
by the green squares, located at the periphery of a ground array of 507
surface detectors (SD).  Each SD unit, shown in figure~\ref{dpf2011_cchjui:fig001} by
the black squares, consists of a scintillation counter mounted on a raised
steel frame.  They are deployed in a square grid of 1.2 km nearest-neighbor
spacing, and the full array covers a total of about 730~km$^{2}$.

Each of the three FD stations has a field of view (FOV) of about
30$^{\circ}$ in elevation and about 110$^{\circ}$ in azimuth.  The central laser
facility (CLF), which initially operated one vertical pulsed YAG laser (355
nm), is marked by the blue cross in figure~\ref{dpf2011_cchjui:fig001}.  The location
of the CLF is equidistant from all three FD stations so that
the vertical pulses can be used for cross-calibration of the three stations
independent of the aerosol concentration in the air.  The three FD stations
are also oriented such that the CLF lies at the center of view of each.

\section{TA Surface Detectors}

Each TA scintillation counter contains two slabs of double-layered plastic
scintillators with an overall collection area of 3.0~m$^{2}$.  The
scintillation photons are collected by wavelength-shifting optical fibers laid
in extruded grooves on the surface of the scintillators.  All of the light
collected from the top and bottom layers are each separately collected into
a single photomultiplier tube (PMT).  Each SD unit is powered entirely by
its own solar panel-battery power supply, and communicates over a 24~GHz
wireless point-to-point link with one of three communication towers.  The
towers are located near each of the three FD stations.

\begin{figure}[ht]
\centerline{
  \includegraphics[width=120mm]{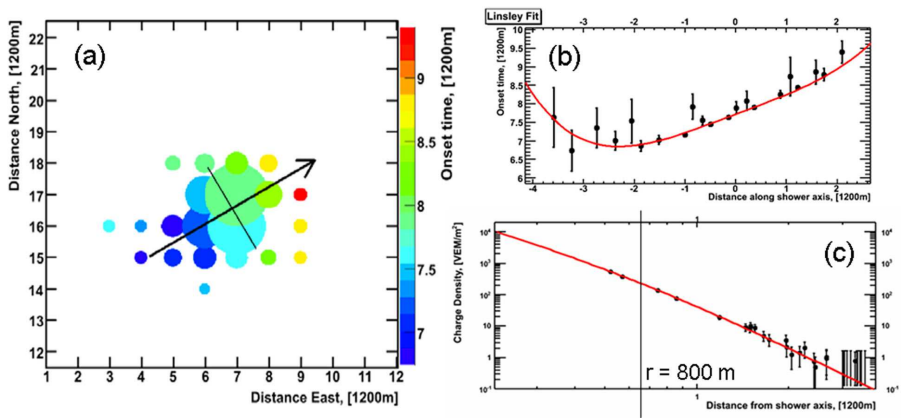}
}
\caption{(a) Left: Display of a typical air shower event captured by
the TA surface detector array, where each circle represents a triggered
SD unit.  (b) Top Right: Fit of onset times of the triggered SD units to 
determine the shower arrival direction. (c) Bottom Right: Fit of the
measured particle density vs. distance from the shower axis and the
interpolation to obtain the density at 800~m (S800).
}
\label{dpf2011_cchjui:fig002}
\end{figure}

The SD counters are self-calibrated using minimum-ionizing cosmic muons. 
these muons provide a convenient unit of "vertical equivalent muon" (VEM). 
The output of the PMT from each scintillator is monitored continuously at
40 million samples per second (40 MSPS).  Pulse data are recorded into a
storage buffer when a cluster of at least 1/3 VEM is observed.  An event
trigger is formed when a minimum of three adjacent counters each detects a
cluster of at least 3 VEM each.  A typical event is shown in
figure~\ref{dpf2011_cchjui:fig002}.

\begin{figure}[ht]
\centerline{
  \includegraphics[width=70mm]{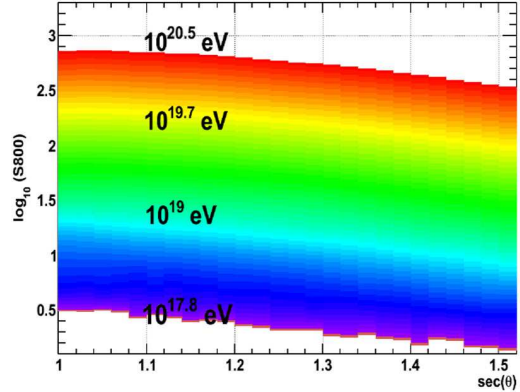}
}
\caption{
The variation of the average particle density at 800 meters
from core (S800) with the energy and zenith angle of simulated air showers.
Energies of real air showers are determined by comparing the measured S800
to this plot.
}
\label{dpf2011_cchjui:fig003}
\end{figure}

In figure~\ref{dpf2011_cchjui:fig002}(a), the hit counters and the their recorded
densities are shown by the circles and their area.  The locations of the
counters are given by row and column number (at 1.2 km spacing).  The color
scheme shows the arrival time in terms of equivalent distance traveled by
light, divided by the detector spacing.  The location, signal size, and
onset time detected by each counter is used to fit for the core (centroid)
location and arrival direction of the shower.  This geometry fit uses a
modified Linsley time delay function \cite{dpf2011_cchjui:linsley} to describe the
curvature of shower front, and the AGASA lateral density function (LDF)
\cite{dpf2011_cchjui:agasa-ldf} to predict the fall of density from the core of the
shower.  The result of this fit is illustrated in
figure~\ref{dpf2011_cchjui:fig002}(b), which plots the onset time vs.  distance
along the direction indicated by the arrow in figure~\ref{dpf2011_cchjui:fig002}(a). 
The particle densities from the hit SD units are then plotted as a function
of its perpendicular distance to the shower core, and the density at 800
meters (S800) is interpolated from the fit to the AGASA LDF function, as
shown in figure~\ref{dpf2011_cchjui:fig002}(c).  The S800 value is then compared to
the average from simulated events (shown in figure~\ref{dpf2011_cchjui:fig003}) and
the measured energy is interpolated according to the measured zenith angle.

\section{TA Fluorescence Detectors}

\begin{figure}[ht]
\centerline{
  \includegraphics[width=120mm]{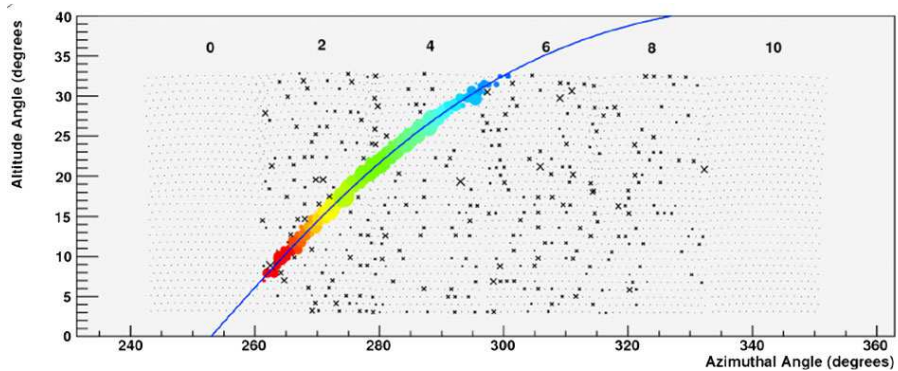}
}
\caption{
Display of a typical downward air shower event captured by the FD station at
Black Rock.  The circles correspond to channels with pulses in
excess of three sigma over the (night sky) background, and the area of the
circles represent the integrated pulse area.  Those pixels associated with
the air shower are easily identified by their size and correlation in
direction and time.  The shower-detector plane (SDP) is shown by the fitted
curve.  The colors indicate arrival time of the signal light, with blue
indicating the earliest and red the latest pulses. 
}
\label{dpf2011_cchjui:fig004}
\end{figure}

A total of 38 fluorescence telescopes are divided into three stations.  The
first of these was constructed on Black Rock (BR) Mesa, at the southeastern
corner of the surface array.  A second station is located at Long Ridge (LR)
on the southwestern flank of the SD array.  The BR and LR detectors were built in
Japan based on essentially the same specifications as the telescopes used by
the HiRes experiment, but with larger mirrors of 6.8~m$^{2}$ area (compared
to 5.2~m$^{2}$ for HiRes).  Each site consists of 12 telescopes with
256-pixel (16x16 in a triangular lattice) cameras.  Each pixel, instrumented
by a hexagonal PMT, covers a cone of 1.1$^{\circ}$ in the sky.  A typical
event captured by the FD station at Black Rock is shown in
figure~\ref{dpf2011_cchjui:fig004}.  For each triggered event, the 10 MHz FADC data
from each channel are scanned for pulses.  Those channels containing a three
sigma excess over background (primarily night sky) fluctuations are displayed
as circles, with the area of the circle being proportional to the integrated
pulse area.  

\begin{figure}[ht]
\centerline{
  \includegraphics[width=120mm]{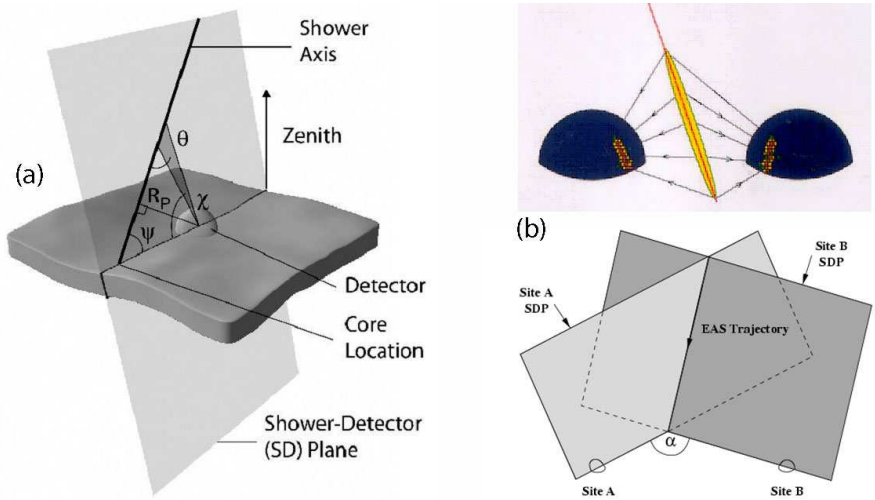}
}
\caption{
(a) Illustration of the monocular FD time fit to determine the
shower trajectory.  (b) Illustration of the intersecting plane method of
finding the shower axis for stereo FD observations. 
}
\label{dpf2011_cchjui:fig005}
\end{figure}

As can be seen in figure~\ref{dpf2011_cchjui:fig004}, the pixels corresponding to
the actual air shower are easily identified by their size as well as spatial
and temporal correlation, and are marked in color.  The pointing directions
of these channels are then used to fit for a shhower-detector plane (SDP). 
Because of the distortion inherent in the Miller cylindrical projection used
for the event display, the fitted SDP appears as a curve in
figure~\ref{dpf2011_cchjui:fig004}.  The colors of event pixels indicate time
progression: blue represents the earliest arrival times at the top of the
event, and red represents the latest at the bottom.  The event depicted was
clearly a downward going air shower.  Once the SDP is obtained, the
trajectory of the air shower can be completely determined in one of two
ways.  For monocular observation, where only the measurement from a single
fluorescence station is used, the shower axis can be determined by fitting
the arrival time at each pixel to equation~\ref{dpf2011_cchjui:eqn_timefit}.
\begin{equation}
t_{i}=t_{0}+\frac{R_{P}}{c}\tan{\left(\frac{\pi-\psi-\chi_{i}}{2}\right)}
\label{dpf2011_cchjui:eqn_timefit}
\end{equation}
As illustrated in figure~\ref{dpf2011_cchjui:fig005}(a), $R_{P}$ is the impact
parameter of the shower (nearest distance of approach of the shower axis to
the FD), and $\psi$ is the angle made by the shower axis to the line of
intersection between the SDP and the ground.  The value $t_{0}$ physically
corresponds to the time at which the shower passes the point of nearest
approach.  The output parameters from the timing fit are $t_{0}$, $R_{P}$,
and $\psi$.  The inputs are the measured times $t_{i}$ and the angles
$\chi_{i}$ of the pixels involved in the event.  As seen in
figure~\ref{dpf2011_cchjui:fig005}(a), $\chi$ is the angle made between the PMT
pointing direction (projected onto the SDP), and the ground, measured within
the shower detector plane.  Alternately, with two FD stations viewing the
same event in stereoscopic mode, the shower trajectory can be determined
from the intersection of the two SDPs.  This stereo reconstruction method is
illustrated in figure~\ref{dpf2011_cchjui:fig005}(b).  Typically at energies in the
ultra high energy (UHE) regime, monocular reconstruction gives $R_{P}$ and
$\psi$ resolutions of about 10\% and $5^{\circ}$, respectively, whereas the
stereo reconstruction improves these to about 5\% and $1^{\circ}$.

\begin{figure}[ht]
\centerline{
  \includegraphics[width=120mm]{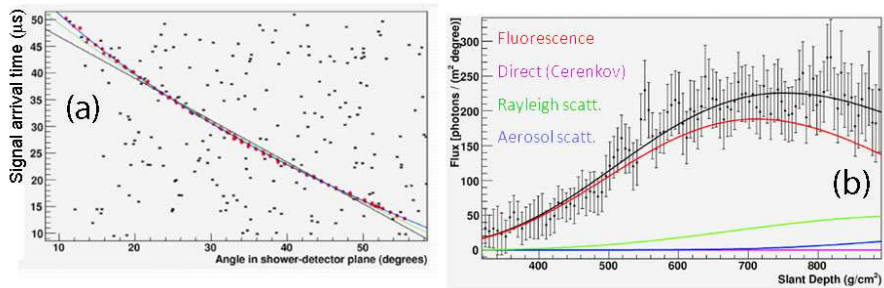}
}
\caption{(a) Timing fit to determine $R_{P}$ and $\psi$. (b) Profile
fit to determine the energy and $X_{max}$ for the shower shown originally
in figure~\ref{dpf2011_cchjui:fig004}.
}
\label{dpf2011_cchjui:fig006}
\end{figure}

Figure~\ref{dpf2011_cchjui:fig006}(a) shows the timing fit described above for the
event shown in figure~\ref{dpf2011_cchjui:fig004}.  The amount of curvature in the
data determines the in-plane angle, $\psi$.  For a given $\psi$, the overall
slope of the data then determines the impact parameter, $R_{P}$.  Having
determined the shower trajectory, the pointing directions of the PMTs are
then converted to slant depth.  The signal is then fitted to a parametric
function, usually the Gaisser-Hillas form \cite{dpf2011_cchjui:gh} for the shower
size vs.  depth, and includes scattered and direct \^{C}erenkov light in
addition to the fluorescence signal.  The profile fit for this same event is
shown in figure~\ref{dpf2011_cchjui:fig006}(b).  The Energy is extracted from the
overall area of the curve, and the depth of the shower maximum, $X_{max}$,
is extracted from the fit.  Over many showers, the $X_{max}$ values give a
statistical measure of the composition of the primary particles.

\begin{figure}[ht]
\centerline{
  \includegraphics[width=50mm]{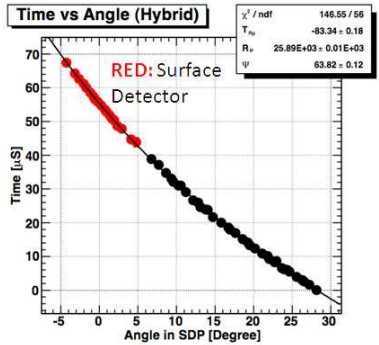}
}
\caption{Timing fit for a hybrid event that includes times of the surface
detectors.
}
\label{dpf2011_cchjui:fig007}
\end{figure}

The High Resolution Fly's Eye experiment used two alternative monocular
reconstruction techniques.  Between 1992-1996, the HiRes prototype, in the
tower configuration (14 telescopes viewing up to $70^{\circ}$ in elevation),
operated in coincidence with the CASA/MIA arrays.  The HiRes/MIA monocular
reconstruction included the timing information from the MIA array.  This
combination became known as the hybrid reconstruction method, and yields
$R_{P}$ and $\psi$ resolutions comparable to that of stereo reconstruction.
The Telescope Array experiment is primarily designed for hybrid FD
reconstruction.  An example of the hybrid timing fit for a TA event is shown
in figure~\ref{dpf2011_cchjui:fig007}. 

Previously for the HiRes-1 site, where the telescopes only view up to
17$^{\circ}$, the observed tracks from air showers are too short for the
timing fit alone to give reliable results.  Instead, another variant of the
monocular reconstruction was used that combined the timing and profile fits
of figure~\ref{dpf2011_cchjui:fig006}.  This technique uses the form of the shower
profile to constrain the range of geometries.  The profile-constrained fit
(PCF) gives resolutions that are comparable to monocular fit at the highest
energies for the one-layer HiRes-1 detector but quickly degrades and becomes
unusable below about $3\times{10}^{18}$~eV.  The third fluorescence detector
station on Middle Drum (MD) Mountain, located at the northern end of TA, was
built with 14 refurbished telescopes from HiRes-1.  This commonality between
HiRes and TA allows us to compare the results of the two experiments
directly.  For this purpose, the initial analysis of the MD monocular FD
data used exactly the same simulation and reconstruction codes as was used
for HiRes-1, changing only the pointing geometry of the detectors, and
lowering the trigger threshold in the simulation to reflect the reduced
ambient background light.

\section{TA Energy Spectrum Measurement}

\begin{figure}[ht]
\centerline{
  \includegraphics[width=80mm]{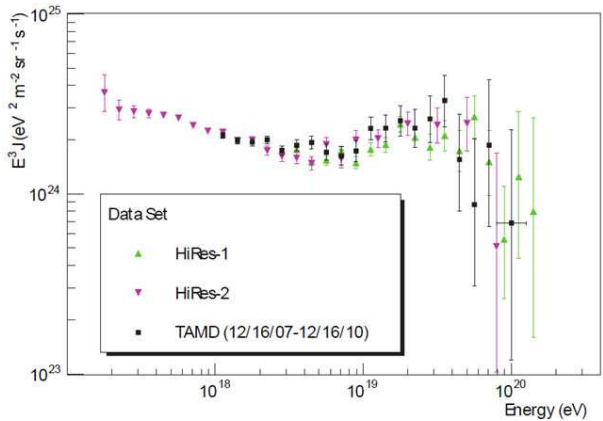}
}
\caption{
TA monocular spectrum from the Middle Drum FD station from its first three
years of observation, overlaid with the monocular spectra from HiRes.  The
TA and HiRes spectra are in excellent agreement.  The TA spectrum
is also consistent with the presence of the GZK cut-off.
}
\label{dpf2011_cchjui:fig008}
\end{figure}

One of the early objectives for building the Telescope Array experiment was
to resolve the discrepancy between the observation of the
Greisen-Zatsepin-K'uzmin \cite{dpf2011_cchjui:gzk} cut-off in the UHE cosmic ray
spectrum.  Using the fluorescence technique alone, HiRes reported the first
observation of the GZK cut-off in 2008 \cite{dpf2011_cchjui:hires_prl2008}, whereas
earlier measurements by AGASA, using a scintillation ground array alone
reported a continuing spectrum \cite{dpf2011_cchjui:agasa_spectrum}. 
Figure~\ref{dpf2011_cchjui:fig008} shows the monocular spectrum from the Middle Drum
FD station from its first three years of observation.  The HiRes monocular
spectrum~\cite{dpf2011_cchjui:hires_prl2008} is also shown in the figure.  The MD
station uses 14 refurbished telescopes from HiRes-1, the latter having
provided most of the statistical significance for the GZK cut-off.  The two
sets of spectra are in excellent agreement both in the shape and overall
normalization.  The new TA result is also consistent with a flux suppression
at the expected GZK threshold.

\begin{figure}[ht]
\centerline{
  \includegraphics[width=100mm]{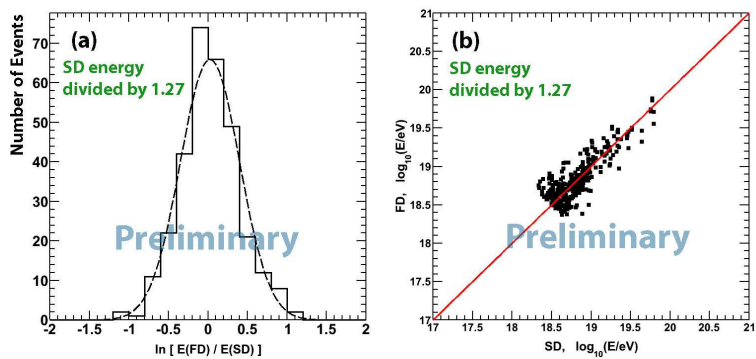}
}
\caption{
(a) A histogram of the difference between FD and SD energies for hybrid
events above $3\times{10}^{18}$~eV.  (b) A scatter-plot of the log FD energy
vs. log SD energy for the same hybrid events.
}
\label{dpf2011_cchjui:fig009}
\end{figure}

From a compilation of TA hybrid events seen by both the SD and the FD, the
SD was seen to give a consistently higher energy.  After the first year of
observation, the SD energies was determined to be consistently 1.27 times
higher than the FD energies. Figure~\ref{dpf2011_cchjui:fig009}(a) shows a histogram
of the difference between the FD and SD energies, with the SD energies
scaled down by 1.27.  A scatter-plot of log FD energy vs log SD energy for
the same events is shown in figure~\ref{dpf2011_cchjui:fig009}(b).  The latter
shows a linear relationship between the two energy measurements over the 1.5
decades of energy above $3\times{10}^{18}$~eV.

\begin{figure}[ht]
\centerline{
  \includegraphics[width=140mm]{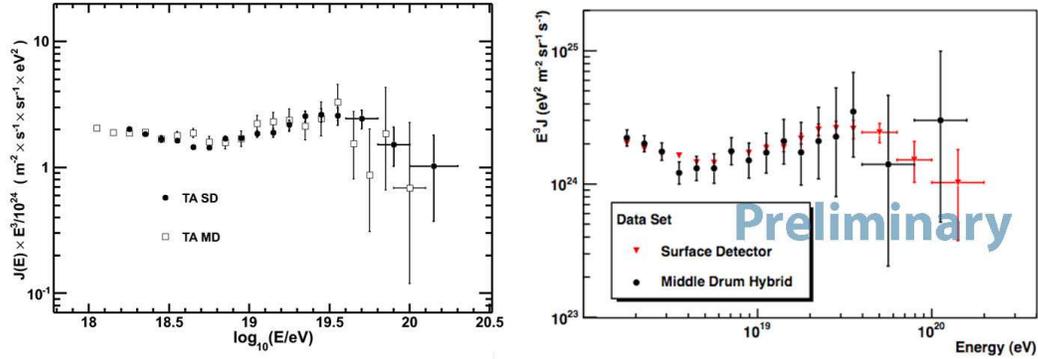}
}
\caption{(a) The TA surface detector spectrum with event energies scaled
down by a factor of 1.27, overlaid with the monocular FD spectrum from
Middle Drum.  (b) The hybrid FD spectrum from Middle Drum overlaid with the
energy-rescaled SD spectrum. 
}
\label{dpf2011_cchjui:fig010}
\end{figure}

Figure~\ref{dpf2011_cchjui:fig010}(a) shows the energy spectrum of UHE cosmic rays
compiled from the first three years of TA surface array data.  The energy of
each event was rescaled by the factor of 1.27.  This spectrum is
overlaid with the monocular FD spectrum (previously shown in
\ref{dpf2011_cchjui:fig008}).  With the rescaling of energies alone, the SD spectrum
obtained is in excellent agreement with the monocular FD spectrum, and in
turn, with the HiRes spectrum both in normalization and in shape. 
Figure~\ref{dpf2011_cchjui:fig010}(b) shows a preliminary hybrid spectrum from the
Middle Drum FD data overlaid with the SD spectrum.  Again the two are in
excellent agreement. We have divided the three TA spectra shown (MD FD
monocular, SD, and MD FD hybrid) and the HiRes spectrum into three plots in
order to avoid clutter.  The conclusion we draw here is that with a 1.27
energy scaling factor for the SD, the TA SD and MD FD spectra are completely
consistent with the HiRes results.  Monocular and hybrid spectra
from BR and LR, not shown here, are also consistent with those of the SD,
and MD.

\section{Composition and Anisotropy}

\begin{figure}[ht]
\centerline{
  \includegraphics[width=140mm]{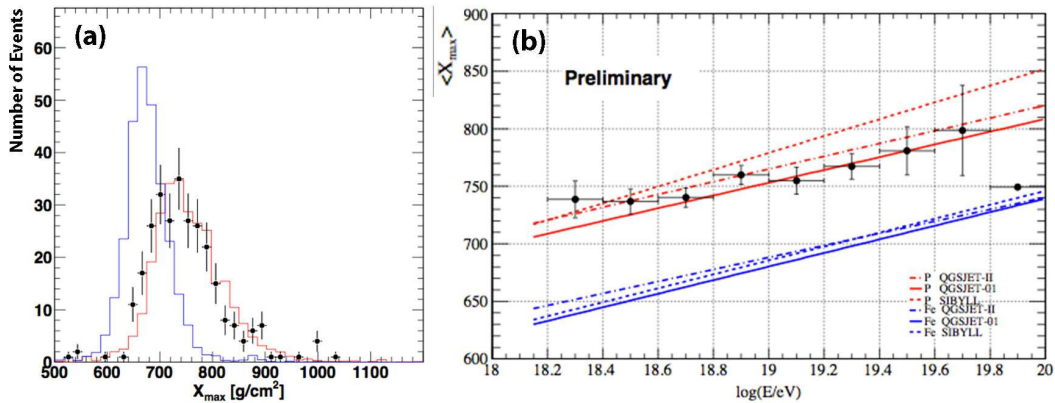}
}
\caption{(a) Distribution of shower maximum depth ($X_{max}$) of TA stereo
data compared to CORSIKA simulation for proton and iron, based on the
QGSJET-II hadronic interaction model.  (b) Plot of mean $X_{max}$ vs log
energy for the same TA stereo data set.  The accompanying curves show
CORSIKA simulation results, including detector response and trigger
selection effects for proton and iron with three different hadronic
interaction models.
}
\label{dpf2011_cchjui:fig011}
\end{figure}

Since 2009, there has been a discrepancy in the $X_{max}$-based composition
results between the AUGER and HiRes collaborations.  AUGER claims to see a
trend toward heavier composition at above ${10}^{19}$~eV
\cite{dpf2011_cchjui:auger_comp}, whereas HiRes results are consistent with a
predominantly proton composition \cite{dpf2011_cchjui:hires_comp}. 
Figure~\ref{dpf2011_cchjui:fig011} shows the first TA composition result based on
$X_{max}$ from stereo events.  In Figure~\ref{dpf2011_cchjui:fig011}(a), the
distribution of $X_{max}$ for TA stereo events is compared to those of iron
and proton events simulated with CORSIKA using the QGSJET-II hadronic model. 
It is clear that in mean value and in width of the distribution, the TA
results are consistent with a predominantly protonic composition. 

Figure~\ref{dpf2011_cchjui:fig011}(b) shows the plot of mean $X_{max}$ vs. log energy
for the same stereo data set.  The various curves show the predictions
(folding in detector response and trigger selection) of CORSIKA simulations
with different hadronic interaction models.  The TA data, like that for
HiRes, is again consistent with a predominately protonic composition,
especially when compared to QGSJET models.  Composition studies based on
the width of the $X_{max}$ distributions, and on the width of the shower
profiles as well as those using hybrid events are nearing completion. 

The anisotropy searches in TA are based primarily on the SD data. After the
first three years of observations, the data is entirely consistent with
isotropy.  We did check the TA data against the claim made by the AUGER
collaboration in the 2007 Science article \cite{dpf2011_cchjui:auger_2007science},
where 8 of 13 AUGER events above $5.7\times{10}^{19}$~eV were seen to be
within 3.1$^{\circ}$ of Active galactic nuclei in the Veron-Cetty
catalog~\cite{dpf2011_cchjui:veron-cetty} with $z<0.018$.  For the northern sky, the
corresponding prediction for TA would have been 15 correlations out of 20 TA
events seen above $5.7\times{10}^{19}$~eV, whereas an isotropic distribution
predicts five accidental correlations.  Of these 20 events, eight were seen
to be in coincidence with AGNs.  This result is not a particularly
significant departure ($p=0.13$) from the null hypothesis. More TA data
is needed for further anisotropy searches.


\begin{acknowledgments}
The Telescope Array experiment is supported 
by the Japan Society for the Promotion of Science through
Grants-in-Aid for Scientific Research on Specially Promoted Research
(21000002) 
``Extreme Phenomena in the Universe Explored by Highest Energy Cosmic
Rays'', 
and the Inter-University Research Program of the Institute for Cosmic Ray 
Research;
by the U.S. National Science Foundation awards PHY-0307098, 
PHY-0601915, PHY-0703893, PHY-0758342, and PHY-0848320 (Utah) and 
PHY-0649681 (Rutgers); 
by the National Research Foundation of Korea 
(2006-0050031, 2007-0056005, 2007-0093860, 2010-0011378, 2010-0028071,
R32-10130);
by the Russian Academy of Sciences, RFBR
grants 10-02-01406a and 11-02-01528a (INR),
IISN project No. 4.4509.10 and 
Belgian Science Policy under IUAP VI/11 (ULB).
The foundations of Dr. Ezekiel R. and Edna Wattis Dumke,
Willard L. Eccles and the George S. and Dolores Dore Eccles
all helped with generous donations. 
The State of Utah supported the project through its Economic Development
Board, and the University of Utah through the 
Office of the Vice President for Research. 
The experimental site became available through the cooperation of the 
Utah School and Institutional Trust Lands Administration (SITLA), 
U.S.~Bureau of Land Management and the U.S.~Air Force. 
We also wish to thank the people and the officials of Millard County,
Utah, for their steadfast and warm support. 
We gratefully acknowledge the contributions from the technical staffs of our
home institutions and the University of Utah Center for High Performance
Computing
(CHPC). 
\end{acknowledgments}


\bigskip 

\end{document}